\documentclass[12pt,preprint]{aastex}

\slugcomment{Accepted by The Astronomical Journal}

\shorttitle{Oscillations of Red Giant Stars}
\shortauthors{Gilliland}

\begin{document}

\title{Photometric Oscillations of Low Luminosity Red Giant Stars\altaffilmark{1}}

\author{Ronald L. Gilliland}
\affil{Space Telescope Science Institute, 3700 San Martin Drive,
Baltimore, MD 21218}
\email{gillil@stsci.edu}

\altaffiltext{1}{Based on observations with the NASA/ESA {\em Hubble
Space Telescope}, obtained at the Space Telescope Science Institute,
which is operated by AURA, Inc., under NASA contract NAS 5-26555.}

\begin{abstract}
I present details of the variations of several hundred red giant stars on time scales of a few hours to a few days from {\em Hubble Space Telescope (HST)}
observations of a low-extinction galactic bulge sample from an intensive
seven day campaign.  Variations in the red giants are shown to be a strong function of position within the color-magnitude diagram (CMD) in accord with general expectations from theory. Amplitudes are greater for stars with larger radii, whether this results from higher luminosity at the same effective temperature or lower temperature at a fixed apparent magnitude.  Likewise, characteristic time scales for the variations increase to the upper right in a CMD as does the ratio of amplitudes measured at 606 nm compared to 814 nm.
Characteristic variation time scales are well matched by low-order radial pulsation modes. The effective sample discussed here extends from about two magnitudes above the bulge turnoff at which red giant radii are $\sim$7 $R/R_{\odot}$ at 5,000 $^\circ$K with typical amplitudes of $\sim$0.5 mmag to $\sim$40 $R/R_{\odot}$ at 4,000 $^\circ$K with amplitudes of $\sim$3.5 mmag.
Variability characteristics are quite similar at any given position in the CMD, and at levels in the CMD where oscillations are easily detected nearly all red giants show such. If these variations represent oscillations with sufficient lifetimes to derive accurate mode frequencies more extensive observations, e.g.\ as should soon be provided by the {\em Kepler Mission}, would provide a rich asteroseismic return.
\end{abstract}


\section{Introduction}

Red giants are attractive targets for the study of stellar oscillations. From the observational perspective, the amplitude of stochastically-driven p-mode oscillations is expected to rise with increasing $L/M$ \citep{kje95}. Assuming that coherent oscillations exist with sufficiently long lifetimes to allow studies analogous to those from solar oscillations, asteroseismology of red giants promises probes of great interest for testing the theory of these highly evolved stars with complicated 
interior structures. For the red giant phase, where evolutionary tracks from different
progenitor masses converge in a CMD, asteroseismology can, in principle, provide valuable constraints on the intrinsic parameters of individual stars \citep{ste08,kal08}.

That red giants show characteristic variations is well known. Less clear (and the subject of increasing study) is what the physical nature of these variations are.  A central question for variations in any group of stars is whether these follow from normal modes of oscillation either driven by nonlinear feedbacks such as in RR Lyrae stars, or from stochastic driving as in the Sun.  The common alternative to oscillations would be variations arising from giant cell convection as suggested by \citet{sch75}, see also \citet{dzi01} and \citet{lud06}, in which case their utility as probes of interior structure likely do not exist as with true oscillations. A secondary question is whether the variations, if oscillations, have lifetimes long enough to allow the accurate determination of frequencies required to support asteroseismic interpretations. This paper, while not resolving these questions, will more firmly establish the overall characteristics of red giant variations by utilizing a serendipitous data set that provides detections on several hundred stars within a narrow confine of the CMD.

After a decade in which the promise was recognized \citep{bro94}, but results did not materialize despite concerted efforts, asteroseismology has provided solid and exciting results in recent years manifested by precise measurements of solar-like
oscillations in a number of stars. See \citet{bed07} for a recent general review and references. For red giants in particular a number of radial velocity studies requiring dedicated use of at least moderately large telescopes and excellent spectrographs have provided evidence for oscillations in red giants. \citet{hat94a,hat94b} found 14--54 m s$^{-1}$ variations in $\alpha$~Boo, following up on the \citet{smi87} detection,  and $\beta$ Oph respectively using observations
over typically eight consecutive nights on the 2.1~m McDonald telescope. These velocity amplitudes would correspond to photometric variations of 0.5--2.0 mmag using the \citet{kje95} scaling relation. With observations over 12 nights and precisions better than those successful for the K1~III star $\alpha$ Boo, \citet{hor96} failed to detect evidence of oscillations on any of four G8~III to K2~III
stars studied. More recently \citet{fra02} detected solar-like oscillations in the
G7~III star $\xi$ Hya peaking at about 2~m s$^{-1}$ from use of the CORALIE spectrograph and 1.2-m Swiss telescope (ESO). \citet{der06} report $\sim$3~m s$^{-1}$ oscillations in the G9.5~III star $\epsilon$~Oph. The power spectrum for the $\xi$ Hya radial velocities has maximum peaks corresponding to periods near 0.15 days, and a generally complex structure with evidence for even mode spacing expected of solar-like, coherent oscillations.  The predicted mode lifetimes from \citet{hou02} are 17 days for this star, while \citet{ste06} find a value of about two days from analysis of the extensive \citet{fra02} observations. The latter value, if true and common, could well limit the overall utility of asteroseismology of red giant stars.

Photometric observations have been quite effective for elucidating variations in extreme red giants to supergiants, e.g.\ \citet{kis06} find ubiquitous variations in supergiants from multi-decade visual observations compiled by the AAVSO that are suggestive of stochastically driven oscillations.  The OGLE-II data base with extensive data over three years also provides \citep{kis03} clear detection of 
multiple variability sequences, presumably related to different radial orders of oscillation for thousands of stars below the tip of the red giant branch from the LMC. For these upper red giant branch variables amplitudes are commonly 1--4\% with periods of 15--20 days with excitation mechanisms related to either Mira-like pulsators or stochastically-driven pulsations from convection. Farther down the RGB \citet{edm96} claimed that red giants often showed variations on time scales of $\sim$2--4 days with amplitudes of 0.5--1.5\% for K~giants in 47~Tuc based on a 40-hour U-band sequence obtained with {\em HST}. \citet{jor97} used extensive Str\"{o}mgren photometry of unusually high photometric stability to detect variations in K~giants down to $\sim$0.5\%, and, significantly, to establish that a clear level of variability onset exists as a function of stellar parameters
above which all red giants are variable. \citet{hen00} reported on extensive automatic telescope observations of 187 red giants with sensitivity to about 0.2\% confirming the \citet{jor97} results and further establishing that most red giants
earlier than G2 and later than K2 are variables.  At the 2~mmag precision level available \citet{hen00} found that most red giants over G3--K1 were not variable.
The variables to the red were interpreted as radial pulsations while those in the blue set had characteristic time scales too long for radial pulsations and non-radial g-modes were proposed.

Extensive ground-based campaigns attempting to detect oscillations in red giants over luminosity, radius ranges covered by the current data have largely fallen short of success. \citet{ste07} observed several giants in M67 and reported indications of excess power broadly consistent with expectations. \citet{fra07} observed many giants in the globular cluster M4 and reported no consistent evidence of oscillations.

Given the extreme range of stellar properties pertaining to red giants, it is not surprising that multiple physical mechanisms may be responsible for subsets of these. With deep surface convection zones red giants are expected to show stochastically excited solar-like pulsations \citep{chr83,hou99,sam07}. \citet{xio07} argue that the hotter type~G giants fall within the classical Cepheid--$\delta$~Scuti instability strip with $\kappa$ mechanism driving, and that a second instability strip for the cooler K and M~giants at log($T_\mathrm{eff}$) $<$ 3.7 is driven by coupling between convection and oscillations. The \citet{xio07} result is broadly consistent with the \citet{hen00} observational results in finding that an intermediate temperature domain is pulsationally stable.

The promise of space-based photometry has recently been demonstrated by \citet{ste08} with the detection of power spectrum excess in 11 red giants observed with the {\em WIRE} satellite.  The {\em WIRE} observations of 15--61 days reach noise levels in amplitude spectra to a best of 7~ppm with a median of just over 20~ppm. Even more impressive are results from the {\em MOST} satellite on $\epsilon$~Oph for which a 28~day observation set for this red giant with $L/L_{\odot} \sim60$, $T_\mathrm{eff}\sim4900$ provides clear and unambiguous detection of multiple, equally spaced modes from which strong asteroseismic 
constraints are quoted by \citet{bar07} and \citet{kal08}. Arguments for relatively long mode lifetimes of 10--20 days are encouraging.

The data available for the present analyses were obtained with {\em HST} ACS as a nearly continuous time series of alternating F606W and F814W exposures for seven days in February 2004. These data were collected for the purpose of extrasolar planet detection for Hot Jupiters orbiting upper main sequence stars 
in the bulge \citep{sah06}. The results here will be of particular interest for extending detections to smaller amplitudes than previously possible from ground-based observations, and for a much larger sample than possible with previous space-based results.

Section 2 will provide an introduction to the properties of these {\em HST} data and discuss the techniques used for time series extractions on saturated stars. The general noise properties of stars on the RGB and main-sequence will be compared at comparable brightness levels and used to show in \S3 that variations in the red giants are ubiquitous at some magnitude levels. The distribution of oscillation characteristics over the CMD will be developed in \S4. Section~5 will provide mappings of RGB position on the CMD to physical stellar parameters of mass, radius, luminosity and temperature and comparison of theoretically expected oscillation properties with those observed. Section~5 will also quantify how the variations detected along the galactic bulge RGB scale with $L/M$, compare with theory, and use this to predict expected variations that will be detectable by the {\em Kepler Mission}.

\section{HST Time-Series of Galactic Bulge Stars}

The GO-9750 program from which these time series have been obtained was
executed over 2004 February 23--29 using the ACS/WFC in filters F606W and F814W. The exposure times were always 339 seconds (the minimum providing 
high duty cycles with the WFC) and supported mmag precision near the bulge 
turnoff as needed for detection of 1--2\% transit signals which were the
primary application \citep{sah06}. Use of the two filters was usually equally split during the 105 {\em HST} orbits. The guiding accuracy of {\em HST} is excellent, but it can experience drifts on a timescale of a day comparable to the 0$\farcs$05 scale of the WFC pixels. To avoid potential systematic drifts in the photometry following from slowly  developing pointing errors the observations were dithered each orbit by a  random value intended to scramble offsets in time which would then allow derivation of intensity versus offset dependence and subsequent correction for this.  The dithers plus inherent drift resulted in offsets that densely 
sampled the scale of a single pixel without exceeding this---exactly as desired.
A total of 254 exposures were acquired in F606W and 265 in F814W providing 
an average of about five samples per 96-minute orbit during the 50~minutes 
available for science.

Full difference image analysis \citep[e.g.\ as used previously by][for a comparable WFPC2 project]{gil00} was utilized for the bulk of the time series work.  This involves building up a model for the intensity each pixel would record \citep{gil99} under the assumption it is monitoring a non-variable target as a function of $x, y$ offsets image-to-image and focus changes.  Derivation of the image-to-image offsets and relative focus variations, as well as elimination of cosmic rays is an inherent part of the difference image analysis (DIA). However, DIA only works for unsaturated objects.  Large step function gradients in flux associated with bleeding from saturated targets are not amenable to such modeling, and I reverted to a simple aperture sum in the direct images for the saturated stars.
Apertures were generated that encompassed all of the pixels bled into from the saturation, the most critical factor in maintaining good results, but with an odd shape tracing the maximum extent of bleeding for a given star with a buffer all around of at least one pixel.  Typical aperture sizes ranged from about 50 pixels for mildly saturated stars to over 600 pixels at 5~magnitudes beyond saturation onset.

All analyses were initiated with the pipeline provided {\bf \_flt.fits} files that contained data already corrected for contemporaneous biases and darks, and that had been flat fielded and converted to electrons detected. Using repeated exposures obtained for an {\em HST} calibration program \citet{gil04} showed that the photometric response for the ACS WFC remained perfectly linear (to precisions near the 0.1\% level allowed by the data) up to and well beyond saturation when the GAIN = 2 option had been selected, which allows the device's full well depth to be reached as the limiting factor rather than the analog to digital converter. One consequence of using simple aperture sums for the saturated stars rather than full DIA is that cosmic rays cannot be eliminated at the 
individual pixel level.  Since a typical cosmic ray has 2000 electrons, cosmic rays can be eliminated as several sigma deviations even on pixels near the full well depth of 84,000 electrons with a good DIA model. However, typical saturated stars may have $4\times10^{6}$ electrons for which a 2,000 electron cosmic ray is exactly a 1 $\sigma$ perturbation---not enough to be flagged and eliminated, but given that the typical star will experience an average of about one per exposure, enough to significantly raise the noise level.  Unusually strong cosmic rays can of course still be flagged and eliminated even in the integrated sums.  

Input to the time series production includes the pipeline data and a master star list \citep{sah06} formed from separate, DAOPHOT \citep{ste92} analyses on over-sampled, combined images in both F606W and F814W. For the red giants, shorter 20 second exposures obtained as part of GO-9750 were also analyzed to provide master list photometry and star positions for stars that would be saturated and thus sometimes blended in the standard 339-second images. Photometry is provided in the Vega magnitude {\em HST} system for the ACS F606W (Broad~V) and F814W~(I) system using the prescription and zeropoints as given by \citet{sir05}. Figure~1 shows the CMD for 230,000 stars derived from these deep,
combined ACS images. Figure~2 shows a blow-up of the brightest few magnitudes of the CMD with superposed fiducial boxes that select subsets of stars by magnitude and color.

Relative time series are formed using the summed counts in individual images minus the mean over all images and then normalized by this mean.  Time series values are left in these direct relative units rather than being transformed to magnitudes. A simple measure of time series quality for the full ensemble of 
stars analyzed is provided by plotting the {\em rms}, or standard deviation against magnitude as shown in Figure~3.  For the stars that do not saturate, difference image analysis provides results that closely approximate limiting precisions expected from a combination of Poisson fluctuations on the object and sky, plus readout noise over nominal apertures.  As discussed earlier the saturated stars, those brighter than F606W $\sim$ 18.72, or F814W $\sim$ 17.73, have an
intrinsic noise level higher by about 20\%, consistent with the inability to eliminate most cosmic rays from their photometric sums. After setting the sums over specially tailored apertures that encompass all of the pixels bled into for saturated targets, plus a small buffer, the relative time series are then decorrelated 
(multiple linear regression) with vectors of x, and y offsets over time, a measure of focus changes, the sky background level, a goodness of fit parameter from the difference image analysis steps, and an ensemble sum over 50 bright, non-variable stars.

The distribution of {\em rms} values for the brightest stars in Figure~3 show positive offsets from limiting values from two effects:  the already mentioned inability to remove cosmic rays that leads to a jump near F606W~= 18.7, and intrinsic variability of red giants, especially for those at the brightest magnitudes.
Justifying the last point will be the primary topic of the next section.

\section{Ubiquity of Variations in the Red Giants}

Fortunately for the purposes of establishing the reality of red giant variations in these data, given that the saturated objects clearly have excess noise as evidenced by the jump in {\em rms} distribution in Figure~3, the CMD contains a large number of foreground dwarf stars at the same intensity levels.

Figures 4 through 8 will contrast photometry for one foreground dwarf and one red giant (each chosen at random) from the intensity level boxes corresponding respectively to F606W~= 15.5, 16.0, 16.5, 17.0 and 17.5 as shown in Figure~2.
Each figure has the same form with direct time series plotted in panels for the dwarf and giant separately at the top, and Lomb-Scargle \citep{sca82} periodograms (square root of this calibrated in units of parts per thousand, or if multiplied by 1.085, are mmag for this amplitude spectrum) at the bottom. The time series are continuous over a seven day period except for losing half of each consecutive 96 minutes to Earth occultation, and cover HJD 2453058.52 to 2453065.47. The time series of the foreground dwarfs are usually closely 
approximated as white noise.  By contrast the time series of red giants at the same brightness level show greatly enhanced variations at low temporal frequencies. The amplitude spectra are plotted from 1.65 to 100 $\mu$Hz, or correspondingly periods of seven days to 2.7 hours. Both the dwarf and red giant time series have no additional structure (not shown) out to the Nyquist limit at about 17 minutes except for expected aliases near the {\em HST} orbital frequency of 174 $\mu$Hz. Variations in the red giants are characterized by slow drifts with similar amplitudes in F606W and F814W and cannot usually be represented as a single sinusoid. At 18th magnitude in F606W the red giant oscillations are no longer noticeable from such comparisons in these data.

To assess the generality of detectable oscillations among the red giants studied here it is informative to examine a histogram of {\em rms} levels for one of the magnitude bins of Figure~2.  Figure~9 shows the {\em rms} distribution at F606W~= 16.5 for dwarfs and red giants summed over bins {\bf B} and {\bf C}
of Figure~2 centered on the RGB. The distribution for the dwarfs is strongly peaked for a bin from 0.0008 to 0.0009 with only a modest asymmetry to the high side.  By contrast there are essentially no red giants at this magnitude and color (3 of 161) with an {\em rms} $<$0.0009, and the peak in the distribution function
for red giants is at an {\em rms} within 0.0011 to 0.0013. A lower limit to time series variability (in units of {\em rms}) at any magnitude may be formed by taking the square root of the difference between the mean red giant and mean dwarf variance (square of {\em rms}).  For example, averaged over all the stars
shown in Figure~9 the red giants have an excess {\em rms} relative to the dwarfs of 0.00074 (ignoring one giant and two dwarfs that have large {\em rms} values more than 0.0015 above the median value).  Means for the dwarfs and giants in Figure~9 are 0.00092 and 0.00118 respectively.  If a significant component of the dwarf variability is intrinsic, rather than just noise from the observations and 
analysis, then the variability ascribed to the red giants should be increased accordingly.  At the levels of nearly one part in a thousand being discussed here the values for the dwarfs are likely to be dominated by observational noise rather than intrinsic oscillation.  Therefore the lower limit to red giant variations established in the simple quadrature difference approach is probably a close approximation to actual levels.

\citet{jor97} and \citet{hen00} had argued that most red giants in restricted regions of the CMD down to sensitivity levels of 0.5\% and 0.2\% respectively are variable.  This survey has shown that red giants significantly farther down the red giant branch are typically variable to sensitivity levels reaching a factor of
10 smaller. 

\section{Distribution of Variation Properties Over the CMD}

Simple physical arguments, e.g.\ as developed by \citet{kje95}, lead to predictions (albeit poorly constrained at this point) of how oscillations should vary with stellar properties. In this section I develop the empirical variations for red giant oscillations with position in the CMD for F606W amplitudes, the frequency of peak power, the relative amplitudes of F606W and F814W variations, a measure of 
mode multiplicity, and a measure of the large splitting (expectation of equal separations in frequency between successive p-modes). In the next section I will establish the stellar properties as the same function of position in the CMD used for collation of oscillation properties and briefly compare expectations from theory to the observations of oscillation properties as a function of stellar parameters.

\subsection{F606W amplitudes}

In \S3 I gave a definition for lower limits to red giant oscillation amplitudes measured in units of {\em rms} by forming the quadrature difference of means for giants and dwarfs at the same magnitude and assuming the dwarfs are non-variable. The results of extending this to six magnitude by five temperature
bins defined in Figure~2 spanning the red giant branch is shown in Table~1.  The excess {\em rms} is shown in units of parts-per-thousand. The equivalent amplitude of a single sinusoid yielding this variation level would be $\sqrt 2$ times the {\em rms} excess. Quantities in parentheses in Table~1 provide a measure of the error on the mean for each bin. The table entries are the means of two independent comparisons of {\em rms} values, one based on means evaluated using values clipped to be within the median plus 0.0015, and one simply based on median values for both the main sequence and red giant stars.  Usually the 
values based on medians and means are very close in value.

A number of things are immediately obvious from Table~1:
(1)~At and above F606W~= 17.5 through 15.5, variability is securely detected in all bins  based on differences of variances, even via this rather crude approach.
(2)~Within each column (with two exceptions out of 21 with the miss well within 1-$\sigma$) roughly corresponding to different temperatures, amplitudes increase monotonically to higher luminosity (assuming stars are at roughly equal 
distances as seems secure from the coherence of features such as the red giant clump seen in the CMD).
(3)~At the same magnitude or luminosity, with only one exception out of 20 steps (and there it is within one-half $\sigma$ on mutual error bars), amplitudes increase to lower temperature.

The bins in Table~1 (and throughout this paper) without numerical entries were avoided for two different reasons. At the two faintest bins for column {\bf A} there are a combination of not many stars and developing overlap with the foreground dwarf distribution. At the two faintest bins in column {\bf E} there would be only one red giant per bin. The total number of dwarfs in the six bins over luminosity entering the statistics for Table~1 is 340, the total number of red giants distributed over 26 bins is 1034.

\subsection{Mean frequency of variations}

The distribution of peak power over the CMD is shown in Figure 10 where medians of amplitude spectra in {\em ppt} are given for the bins spanning both 
main sequence (taken as controls) and red giants. For Figure~10 the combined time series in F606W and F814W were analyzed simultaneously. The variation of amplitudes introduced in the previous section and shown in Table~1 are easily 
visualized in Figure~10.  Noteworthy are the nearly flat median amplitude spectra for the main sequence stars with just a hint of turn up (possible 1/{\em f} noise)
below frequencies of 5~$\mu$Hz, or periods greater than two days.

Inspection of Figure 10 strongly suggests that with increasing luminosity or decreasing temperature the location of peak power moves to lower frequency.  I have quantified this by defining a range of frequency in each star spanning the first and last locations with amplitude greater than or equal to half the peak
amplitude up to 70~$\mu$Hz and then computing the mean frequency over the
domain weighted by the amplitude spectrum. Only stars for which the power
peak at frequency less than 70~$\mu$Hz exceed a significance limit of 0.001 for chance of false alarms \citep{sca82} are included in the statistics. Within each row and column of Table~2 the changes of typical frequency at which oscillations peak occurs in the expected way.

\subsection{Relative amplitudes of F606W and F814W variations}

The variation of relative amplitudes with bandpass over the CMD is shown in Table~3. These are computed by Gaussian smoothing and interpolating each of the F606W and F814W time series to a common grid (with full width half maximum of Gaussian at 2~hours), then performing a linear fit of F606W to F814W.  Stars are retained using three cuts: as for Table~2 and mean frequency the star must have a power spectrum peak with a false alarm probability less than 0.001 and the peak must be at less than 70~$\mu$Hz, in addition, the linear correlation
of F606W and F814W filtered time series must be $>$0.6. Here distinct trends are clear with larger amplitudes in F606W relative to F814W at brighter and cooler stars.

\subsection{Mode multiplicity measure}

Most red giants that show any power spectrum peak with a false alarm probability of less than 0.001 show multiple peaks in the power spectrum. To quantify this I have counted the number of peaks below 70~$\mu$Hz at an amplitude greater than four times the mean amplitude spectrum averaged over 70--100~$\mu$Hz.
This algorithm returned 3, 3, 4, 2, 1 peaks for the red giants plotted in Figures~4 through 8 respectively.  None of the dwarf stars plotted in these figures showed any peaks above this mean level. Table~4 shows the distribution of mean number of peaks over the CMD. Generally lower values in the left-most column and faintest row follow from having barely enough S/N to detect oscillations. The number of independent peaks for stars toward the upper right corner of the table are limited by the frequency resolution of these data being insufficient to resolve independent peaks (see the pile up at lowest frequency for these bins in Figure~10).  The only conclusion is that red giants that show variations are characterized by changes on several independent frequencies rather than commonly displaying 
only one or two modes.

\subsection{Large splitting of modes}

A power spectrum of solar oscillations shows a well populated set of modes at nearly equal frequency separation.  The most evident separation is called the ``large splitting" and corresponds to modes that differ by one in radial number, or two in the number of nodes around the stellar circumference.  Measuring the large splitting for stars displaying solar-like oscillations immediately gives a
good constraint on the mean density of the star.

Inspection of Figure 4 is suggestive of three modes at equal separation. I have attempted to quantify the large separation for the 200 red giants that each showed at least four peaks at greater than four times the mean amplitude spectrum averaged over 70--100~$\mu$Hz by computing the power spectrum of the power spectrum over the frequency domain spanning these significant peaks.  Unfortunately for every red giant like the one in Figure~4 with regularity in the power spectrum, there are several more red giants for which the power spectrum of the power spectrum fails to reflect such regularity. A formal evaluation of median large splittings as the location of the highest peak in the power spectrum of the power spectrum up to 15~$\mu$Hz fails to show any trends over the CMD and I have elected not to show the resulting table.

These time series are relatively short for quantifying frequency  characteristics of the oscillations in red giants.  Reliable detections of large splittings for a large ensemble of red giants (see discussion of prospects with {\em Kepler} in \S5.3) awaits collection of more extensive observations.  It remains to be seen whether red giants will typically have mode lifetimes sufficiently long to allow frequent
measurement of the large splitting.

\section{Relation to Stellar Physical Properties}

In the previous section observational results were developed that show clear correlations of variation amplitudes, relative amplitudes in F606W and F814W, and mean frequency of variations for a wide range of galactic bulge red giants over the CMD.  In order to provide interpretation of these measured variation/oscillation properties, and to make predictions for other characteristics for which these data were insufficient, it is necessary to have estimates of stellar
physical parameters for the observed stars.  Lacking any detailed knowledge for individual stars in this field beyond photometry in two {\em HST} ACS bandpasses, I make use of the easy to use and conceptually powerful simulation package BaSTI (for A Bag of Stellar Tracks and Isochrones) documented in
\citet{pie04}, \citet{pie06}, and \citet{cor07}.

In particular I have utilized the BaSTI web tool for generating synthetic color-magnitude diagrams with input appropriate for the galactic bulge population.  The option for simulating the Milky Way Bulge star formation history was selected; this is based on \citet{mol00}.  The synthetic CMD provided by BaSTI included only standard Johnson-Cousins photometric bandpass simulation.  In order to transform to the ACS F606W and F814W system in use for these observations I ran identical isochrone simulations with BaSTI appropriate to the galactic bulge RGB for which output in both Johnson-Cousins and ACS
bandpasses were available, and derived a simple empirical transform as needed to take the CMD simulations to the ACS system. Having input values of 14.3 and 0.64 \citep{sah06} for distance modulus and E(B--V) color excess for the BaSTI CMD simulation, transforming photometric results to the ACS system a further tweak of
0.05 in F606W--F814W provided the simulated photometry shown in Figure~11 which compares well with the observations in Figure~2.  Random errors of 0.1 magnitude in F606W and 0.03 in F606W--F814W were also applied to the BaSTI simulations shown in Figure~11.

\subsection{Mass, radius, luminosity, temperature over the CMD}

For the purposes here of developing a rough understanding of how intrinsic stellar properties vary over this observational data set the matching simulation from BaSTI shown in Figure~11 allows easy derivation of mean physical parameters corresponding to the boxes in F606W versus F606W--F814W used throughout this paper.

The output from BaSTI includes age, [Fe/H], luminosity, mass, effective 
temperature and apparent photometry for a large ensemble of simulated stars.
Having transformed the photometry to the ACS system it is now a simple matter to compile averages within the CMD boxes for stellar physical properties of 
interest.

Table 5 shows mean luminosities in solar units from the BaSTI simulations.  Over the range of F606W~= 17.5 in box~{\bf B} of the CMD through F606W~= 15.5 in box~{\bf E} for which RGB variations are securely indicated for the majority of stars the stellar luminosities range from 15 to 350 solar, or a dynamic range
of more than 20.  

Effective temperature means are shown in the middle block of Table~5 with a range from about 5100~K to 3950~K corresponding to spectral types of about G4 to K3 giants \citep{all73}.

Stellar radii are shown in the lower block of Table~5 with a range from about 
5 to 40 in units of the solar radius.  Not shown, but available from BaSTI are results for mass and intrinsic metallicity.  The mass values vary little, being about 
0.9~$M_{\odot}$ at the highest temperatures to about 1.05~$M_{\odot}$ at the lowest temperatures. Metallicities range from $-1.0$ for box~{\bf A} at the left side
of the RGB to 0.0 for box~{\bf E} at the right side.  The [Fe/H] values are unique in showing relatively large distributions within these CMD boxes, with dispersions within boxes often near a factor of two.  Other quantities like luminosities, temperatures, radii and masses are much more tightly coupled to CMD position with dispersions within boxes remaining small compared to mean changes box-to-box within the CMD.

\subsection{Expected oscillation properties compared to observations}

With values of typical stellar properties of luminosity, radius and effective temperature shown in the previous section the corresponding theoretically expected oscillation properties will be developed here.

For the Sun, p-modes show a broad distribution centered on 3020~$\mu$Hz.  \citet{bro91} argued that the location of maximum power would likely remain constant in ratio to the acoustic cut-off frequency for other stars. This may be expressed as \citep[following][]{kje95}:
\begin{equation}
\nu_{max} = 3021 \frac{M/M_{\odot}}{\left(R/R_{\odot}\right)^2 
\left(T_\mathrm{eff}/5777\right)^{1/2}} \mu Hz
\end{equation}

Using inputs from Table 5 plus mass values differing only slightly from one solar mass (but included) the predicted location of maximum power from Eq.~(1) is shown in the top block of Table~6.  Note that the variation from lower left to upper right in the observational version of this (Table~2) is in good agreement as to simple trends. In detail the theoretical values of Table~6 decline to much
shorter frequencies for the most luminous and cool stars than is seen in the short observation sets available here.

As expected theoretically, but still with considerable uncertainty, the most likely modes of oscillation may be low-order p-modes \citep{dzi01,xio07} in which case the frequencies would be somewhat different than given by Eq.~(1) and Table~6.
\citet{kal05} presented results for modelling a 47~Tuc red giant with $L/L_{\odot} = 119$, $R/R_{\odot} = 18.8$, and $M/M_{\odot} = 0.9$ which very closely matches box~{\bf C} at F606W~= 16.0. With a $<$4\% adjustment for square root of relative density between the \citet{kal05} model and this box I estimate the 3rd radial overtone frequency at 8.42~$\mu$Hz. Scaling this result by $<\!\rho\!> ^{1/2}$ leads to the values shown in the middle block of Table~6.  These frequencies are a considerably better match to the location of power maxima shown in Table~2. Allowing the freedom to select different radial overtones
from the fundamental through 5th as a function of CMD position would allow essentially perfect matches.

The expected large splitting for modes that differ by one in radial order, or two in azimuthal is inversely proportion to the stellar mean density (with correction factors of only a few percent needed).  The modes of even and odd azimuthal
order produce an apparent evenly spaced set of modes separated as:
\begin{equation}
\Delta \nu_o /2 = (M/M_{\odot})^{1/2}\ (R/R_{\odot})^{-3/2}\ 67.45\ \mu Hz
\end{equation}

The lowest block in Table~6 shows the implied separation of equally spaced 
non-radial modes.  The extent of the observations provides a resolution of 1.65 $\mu$Hz, the upper-right half of the table has large splittings that are smaller than the resolution available.  While these observations are useful for broadly illuminating properties such as typical amplitudes, they are clearly not sufficient to pursue this interesting quantity required to perform basic asteroseismology.
At the magnitude-temperature combinations in Table~6 where splittings exceed 3~$\mu$Hz and might be expected to be robustly detectable in these data from a resolution perspective, Table~1 shows that there is very little overlap with respect to clear indications of excess power.

For low-order radial modes typical frequency separations would be 1.5--2 times those shown in Table~6.  Over much of the CMD domain the resolution provided by these data would still be insufficient to separate individual modes. However, stars at F606W~= 17.0--17.5 for boxes~{\bf B} and {\bf C} might be amenable to more detailed study---a topic for future investigation for individually favorable stars.

The remaining lowest order characteristic of oscillations worth comparing theoretical and realized results for is the variation amplitude.  For this the literature indicates significantly less confidence in providing a unique prediction.  I have adopted two predictions in units of parts-per-thousand from \citet{kje95} and \citet{sam07} with equations of:
\begin{equation}
(\delta L/L) = 4.27 \times 10^{-3} \frac{L/L_{\odot}}{\left(T_\mathrm{eff}/5777\right)^2 M/M_{\odot}}
\end{equation}

\begin{equation}
(\delta L/L) = 4.27 \times 10^{-3}\ (L/M)^{0.7}
\end{equation}

The results from Eq.~(3) of \citet{kje95} are shown in the upper block of Table~7, while those from Eq.~(4) from \citet{sam07} are given in the lower Table~7 block. Comparing these two tables to Table~1, which shows the excess {\em rms}
of the observed time series, demonstrates excellent agreement in terms of trends over the CMD.  Taking into account that {\em rms} excess is not the same to within factors of a few as the theoretically given oscillation amplitudes, it is not possible to strongly discriminate between the two theoretical predictions. Eq.~(3) generally suggests larger amplitudes than exist in the data. Eq.~(4) lacks the strong dependence on $T_\mathrm{eff}$ that seems to be present in the data, and has amplitudes that are too small. Furthermore both theoretical predictions show a relatively larger spread within columns over the magnitude range than 
is seen in the data, e.g.\ for column~D over 17.5 to 15.5 the observations imply a factor of 4.0 change while the two theoretical evaluations yield 12.8 and 5.0.  In the latter regard the simpler $L/M$ scaling of the \citet{sam07} relation is much closer to these observations, and the existence of many independent modes summing incoherently can explain at least part of the offset between absolute values.

The general rise of amplitudes to lower effective temperature in these data is broadly consistent with observational claims of \citet{hen00} for generally more luminous red giants, and with theory as in \citet{xio07}. The distribution of red giants in this study does not extend to early G where amplitudes might be expected to rise to higher effective temperatures.

The observations here for which excess power has been detected in most time series span a range of 20 in $L/M$, and it will therefore be informative to attempt to develop fits for the amplitude as a function of stellar parameters from these data.
I ran experiments performing linear regressions of the {\em rms} excesses of Table~1 against logarithmic combinations of luminosity, mass, temperature and radius in various combinations. The reduced $\chi ^2$ versus a fit in just $L/M$ is 7.74---consistent with the large scatter in the left panel of Figure~12. From inspection of Figure 12 the amplitude excesses show systematic offsets corresponding to left-right position within the red giant branch.  Since left-right position is associated with strong temperature changes a regression in $L/M$ and 
$T_\mathrm{eff}/5777$ was performed dropping the reduced $\chi ^2$ to 1.68 as shown in the right panel of Figure~12. This formal fit is:
\begin{equation}
\mathrm{Amplitude} = -1.34 + 0.483 \log _{10}(L/M) -14.2 \log _{10}(T_\mathrm{eff}/5777)
\end{equation}

Since the amplitude fit here is only a rough indication of excess variation from all sources, not just potential coherent oscillations as might be revealed by longer and more precise observations, general adoption of this equation is not intended to represent a prediction of coherent oscillation amplitudes in red giants.

\subsection{Significance for Kepler Mission}

These observations have been powerful in indicating the ubiquity of variations with characteristic time scales similar to those expected from stellar oscillations in nearly all red giant stars observed with luminosities above 20--30~$L/L_{\odot}$.
In terms of providing results of direct asteroseismic significance, e.g.\ through determination of the large splitting for solar-like p-mode oscillations which would yield a constraint on stellar density, these results are severely limited by both the precision and frequency resolution which are generally inadequate to provide
such results.

The {\em Kepler Mission} \citep{bor07} which has a primary goal of detecting
Earth-like planets transiting roughly solar-like stars also carries a goal of characterizing the host stars in order to better determine planet parameters and the context in which they are formed and exist.  Two key tools to be used by {\em Kepler} in characterizing a large subset of the 100,000 stars to be observed nearly continuously for 3.5 years will be the derivation of astrometric parallaxes, and asteroseismic probing of interiors. Ninety-nine percent of the stars to be observed will be chosen on the basis of yielding the best possible chances for detecting small orbiting planets, thus restricting such choices to dwarfs. However, current plans call for observing 1,000 red giants at the standard 30-minute cadence for 3.5 years for the dual purposes of:  (a)~providing low parallax controls, and (b)~asteroseismic results expected to much better constrain our
knowledge of these stars and thus contribute to a better overall understanding of stellar characteristics within the {\em Kepler} field.  In order to serve as astrometric standards it is important that targets not be allowed to saturate in the
individual 3--8 second ($\sim$6 seconds currently favored) exposures which will be co-added to 30 minutes on board.  This sets a  magnitude limit of about 11.7.  A test selection of 1,000 red giants distributed over the focal plane chosen to never
saturate, weighted to be at large distance, and generally lacking in close neighbors that would compromise astrometry have a mean magnitude of 11.98 (in a system appropriate to the broad bandpass of {\em Kepler} from 400 to 850 nm).  At this magnitude the expected precision will be 65 ppm per 30 minute sum.
After a one-week period the {\em Kepler} photometry should yield power spectrum noise levels of 3.5 ppm for these 1,000 red giants compared to a {\em best} noise level of about 35 ppm in these {\em HST} observations of one week.  More importantly the {\em Kepler} observations are expected to be nearly 
continuous for a full 3.5 years.  After the acquisition of three months of data the precisions provided for stars near 12th magnitude should reach about 1.0 ppm  
(sufficient to detect oscillations in solar-like stars, let alone the red giant variations at much larger amplitudes), and provide a frequency resolution adequate to critically sample the evenly spaced p-mode distribution for stars corresponding to all of the CMD positions detailed in this paper.  Out of the 1000 red giants selected on criteria primarily intended to allow excellent astrometric measures from the {\em Kepler} data, application of Eq.~(5) yields only 31 stars that have predicted {\em rms} excesses below 100 ppm; the majority are expected to have excesses greater than 1000 ppm.

With a launch expected in February 2009 {\em Kepler} should quickly provide a vast wealth of information bringing clarity to properties of oscillations in red giants that have only been weakly explored to date.

\section{Summary}

Observations over a seven day span with the ACS imager on {\em HST} have been used to demonstrate that most red giants with luminosities between 30 and 350~$L_{\odot}$  show evidence of excess noise in power spectra in comparison to main sequence control stars at the same brightness. Amplitudes and frequencies depend upon position in the CMD, and thus differences of stellar parameters of $L$, $M$ and $T_\mathrm{eff}$ in rough agreement with theoretical expectations.  The best match of frequency for observed power excess in the red giants is consistent with interpretation as low-order radial modes.
Since the frequency resolution of these data are not sufficient to resolve expected
oscillation characteristics over much of the sampled domain, asteroseismic applications are not provided.  The upcoming {\em Kepler Mission} will provide improvements of more than an order of magnitude in both photometric precisions and frequency resolution allowing definitive understanding of the nature of
variations in red giants all along the branch.

\acknowledgments

I thank Kailash Sahu for initiating the GO-9750 program and for numerous discussions regarding data analyses.  I also thank Tim Brown, Hans Kjeldsen, Dimitar Sasselov and Dennis Stello for discussion on the topic of red giant oscillations. Daniel Cordier, Adriano Pietrinferni, and Santi Cassisi provided 
correspondence and access to the easy to use and powerful BaSTI simulation package for which I am very grateful. Access to an early release version of the {\em Kepler Input Catalog} was kindly provided by Dave Latham, Tim Brown and Dave Monet. The anonymous referee provided useful comments on the paper.
Financial support for this work was provided through programs GO-9750 and GO/AR-11254 from STScI.

{\it Facilities:} \facility{HST (ACS)}.

\clearpage

\begin{figure}
\plotone{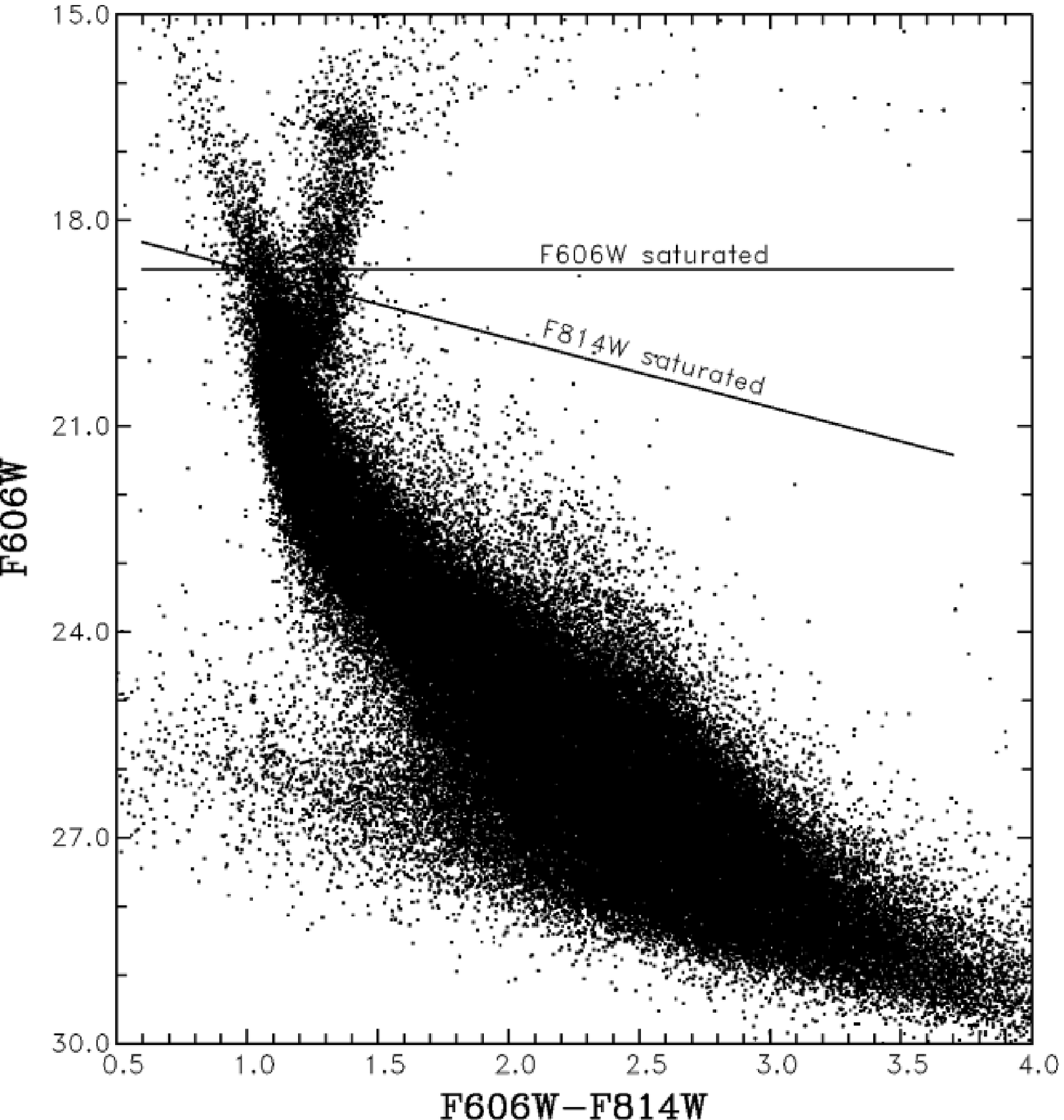}
\caption{CMD for 230,000 stars identified from combined, over-sampled images consisting of 86,106 (F606W) and 89,835 (F814W) seconds with ACS on the galactic bulge.
\label{fig1}}
\end{figure}

\begin{figure}
\plotone{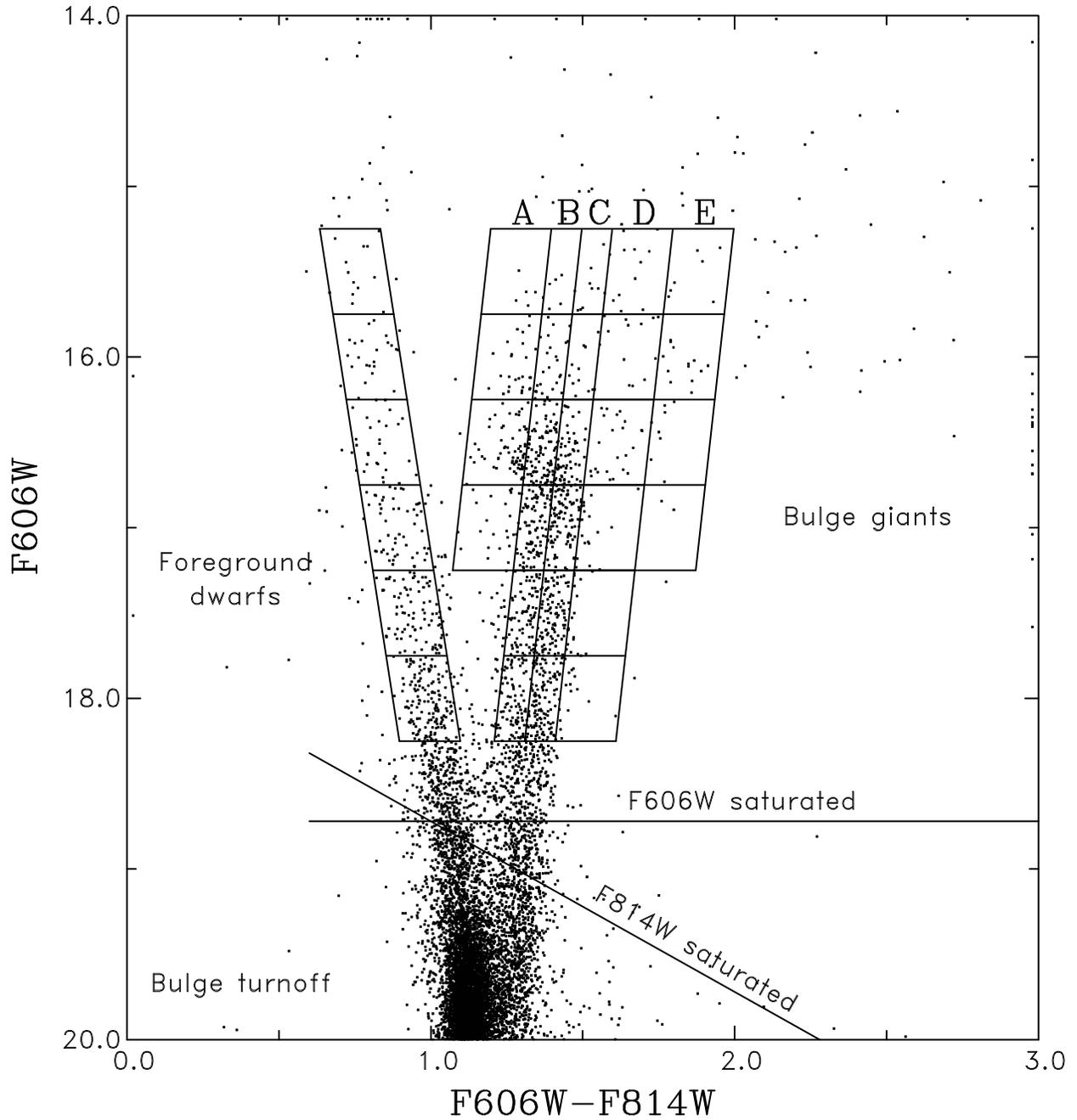}
\caption{Region of CMD from Figure 1 replotted on a larger scale to emphasize the red giant domain.  The over-plotted grids will be referred to throughout the paper to demonstrate changing properties of variations as a function of stellar parameters.
\label{fig2}}
\end{figure}

\begin{figure}
\plotone{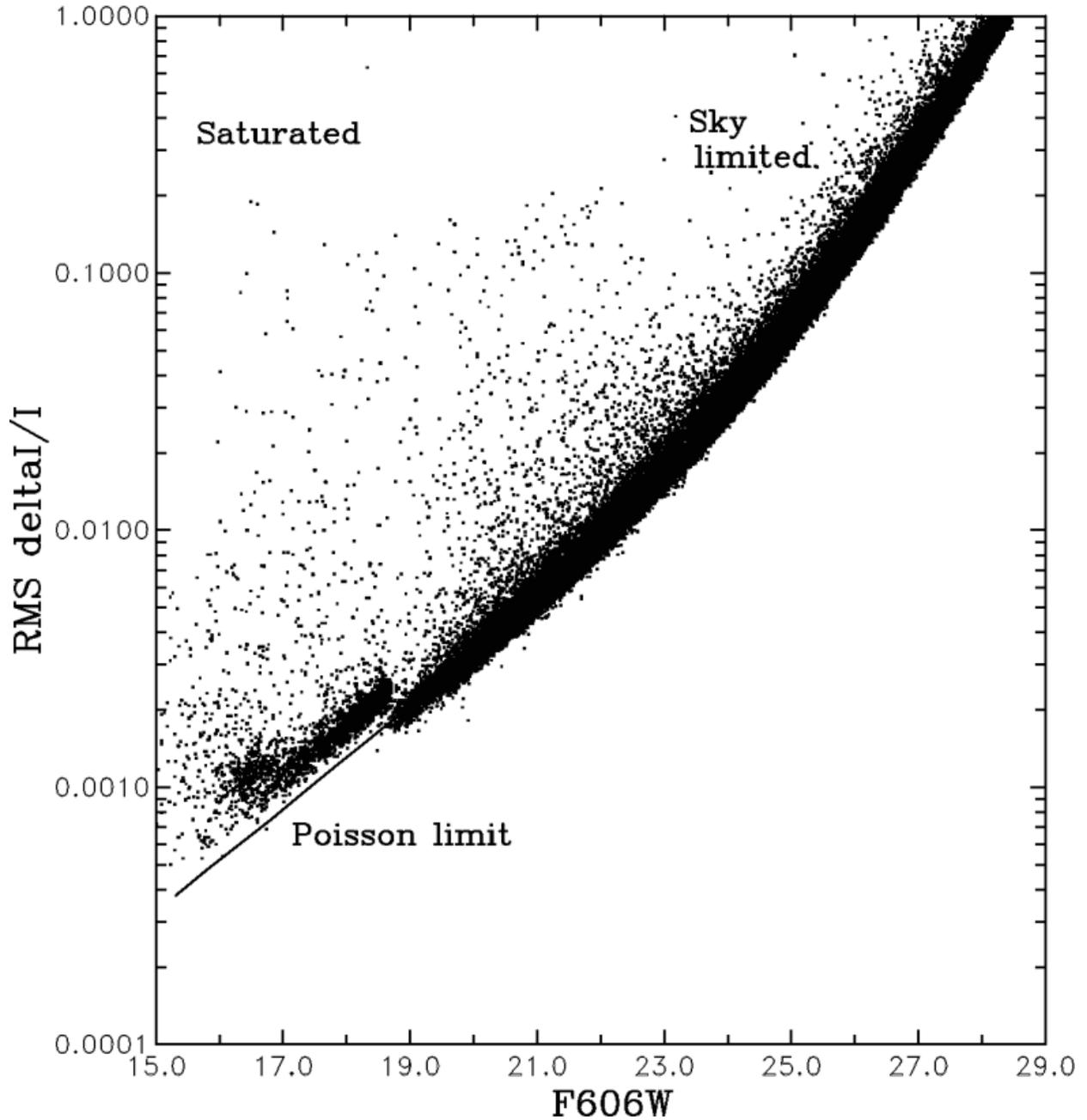}
\caption{Shows standard deviation of relative time series intensity for all 230,000 stars shown in Figure~1.  The break near F606W~= 18.5 results from transition from difference image analysis (fainter) to simple aperture photometry as discussed in the text.  The solid curve shows expected limiting precision (Poisson on objects and sky, plus readout noise) which closely tracks results for the non-saturated star photometry.
\label{fig3}}
\end{figure}

\begin{figure}
\plotone{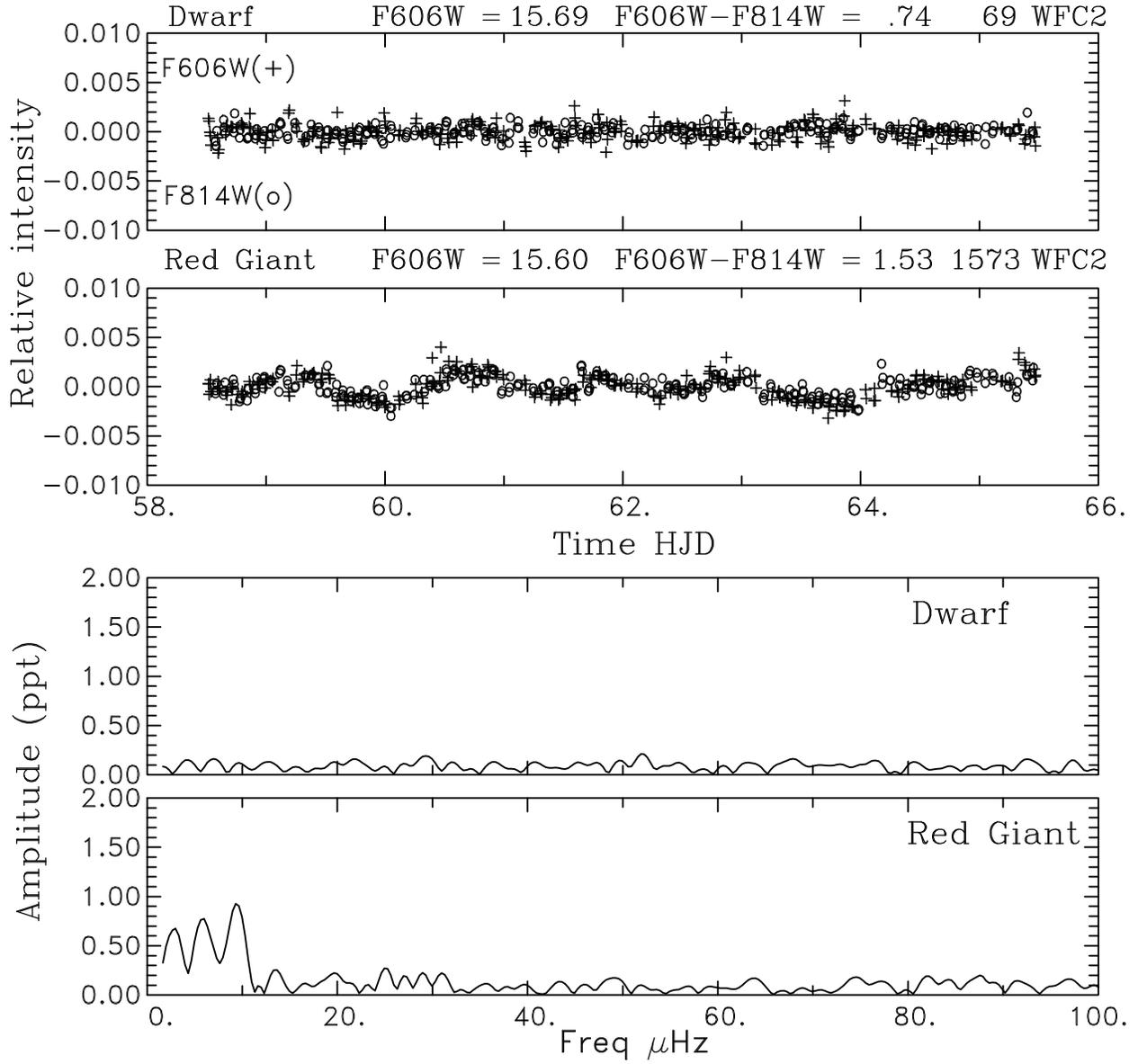}
\caption{Compares time series for a foreground dwarf and red giant star at comparable magnitude of F606W $\sim$15.5.  Top two panels for the dwarf and red giant respectively show the direct, relative photometry spanning 7~days with results for both filters shown. The bottom two panels show square root of the Lomb-Scargle periodogram \citep{sca82} calibrated to units of parts-per-thousand.  Note that the red giant shows excess variability over a domain of 2.5--10~$\mu$Hz (about 1--5 days).
\label{fig4}}
\end{figure}

\begin{figure}
\plotone{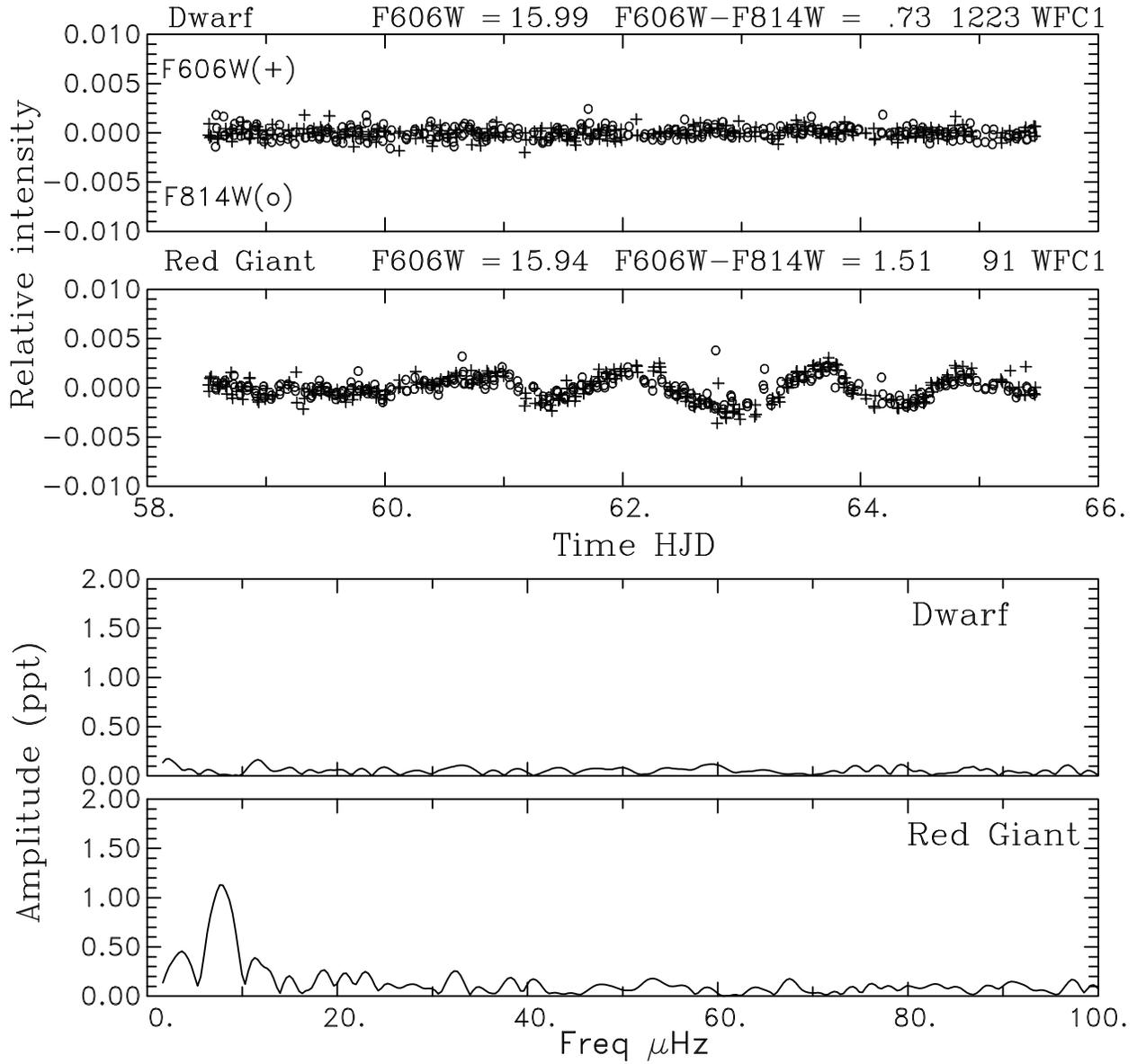}
\caption{Same as Figure 4, but for F606W $\sim$16.0.
\label{fig5}}
\end{figure}

\begin{figure}
\plotone{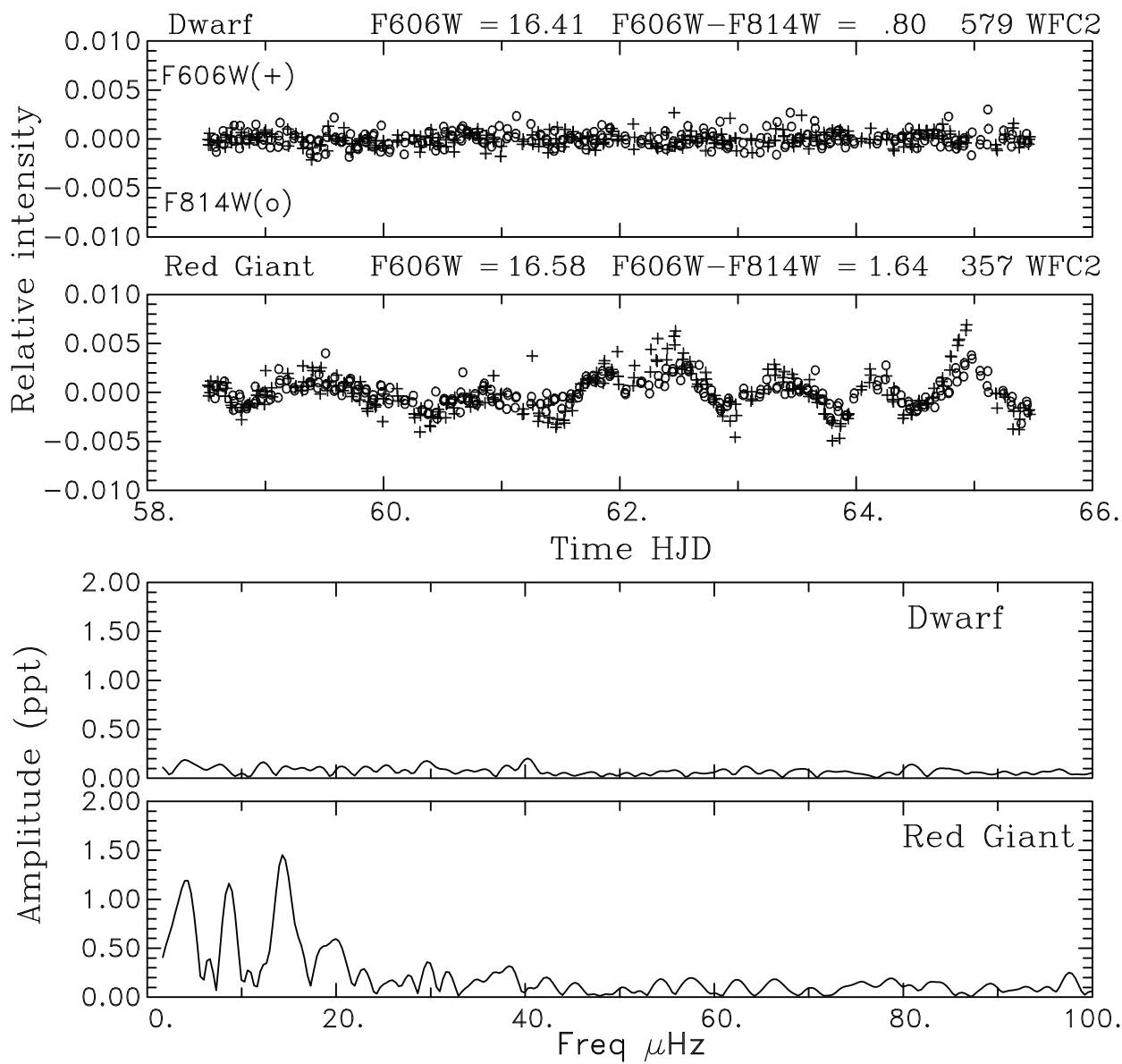}
\caption{Same as Figure~4, but for F606W $\sim$16.5. Excess variability extends dows to periods of about one-half day.
\label{fig6}}
\end{figure}

\begin{figure}
\plotone{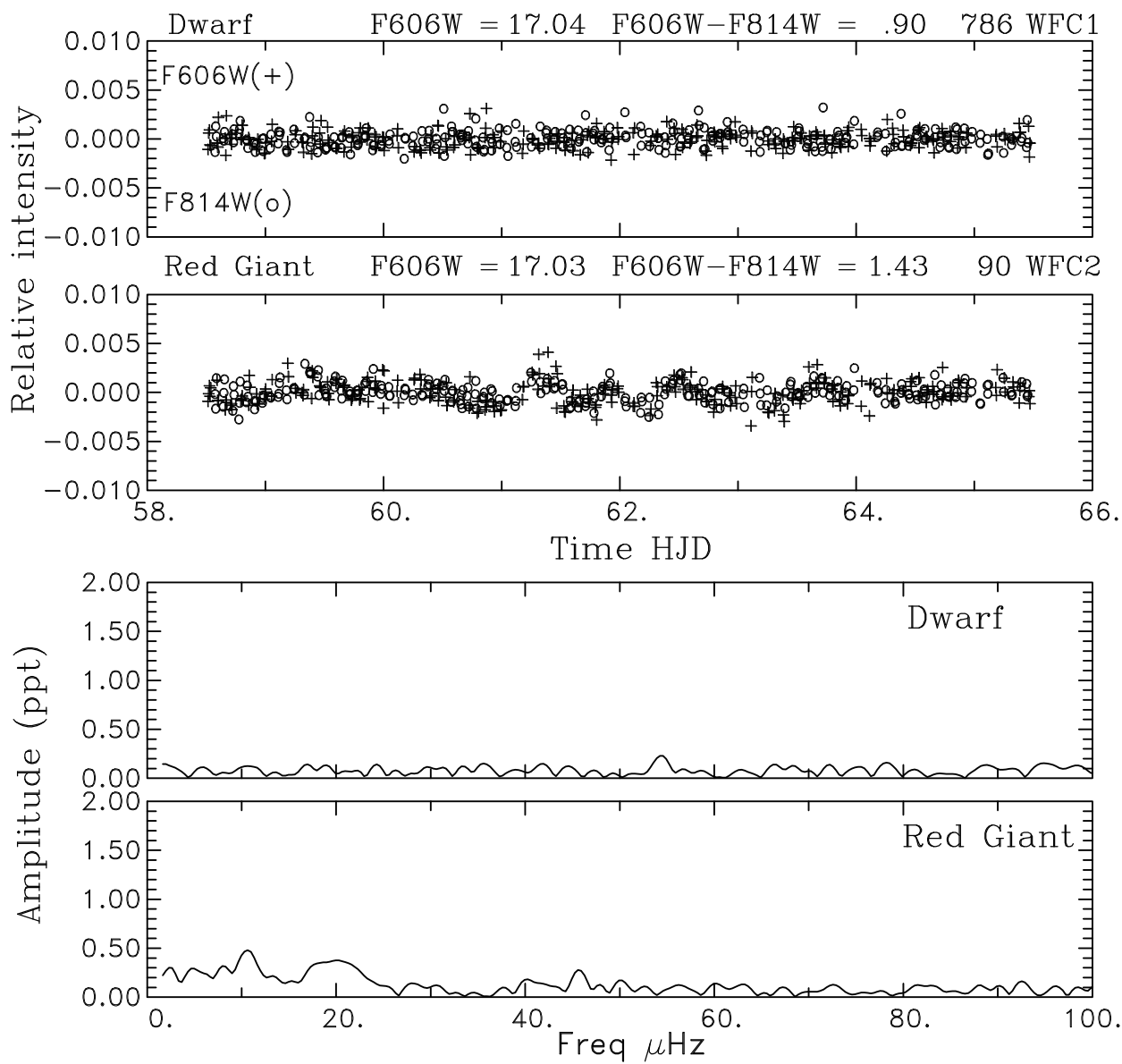}
\caption{Same as Figure 4, but for F606W $\sim$17.0. Although excess variability is still nocticeable in the red giant, the amplitude is considerably smaller than is typical of giants at higher luminosity.
\label{fig7}}
\end{figure}

\begin{figure}
\plotone{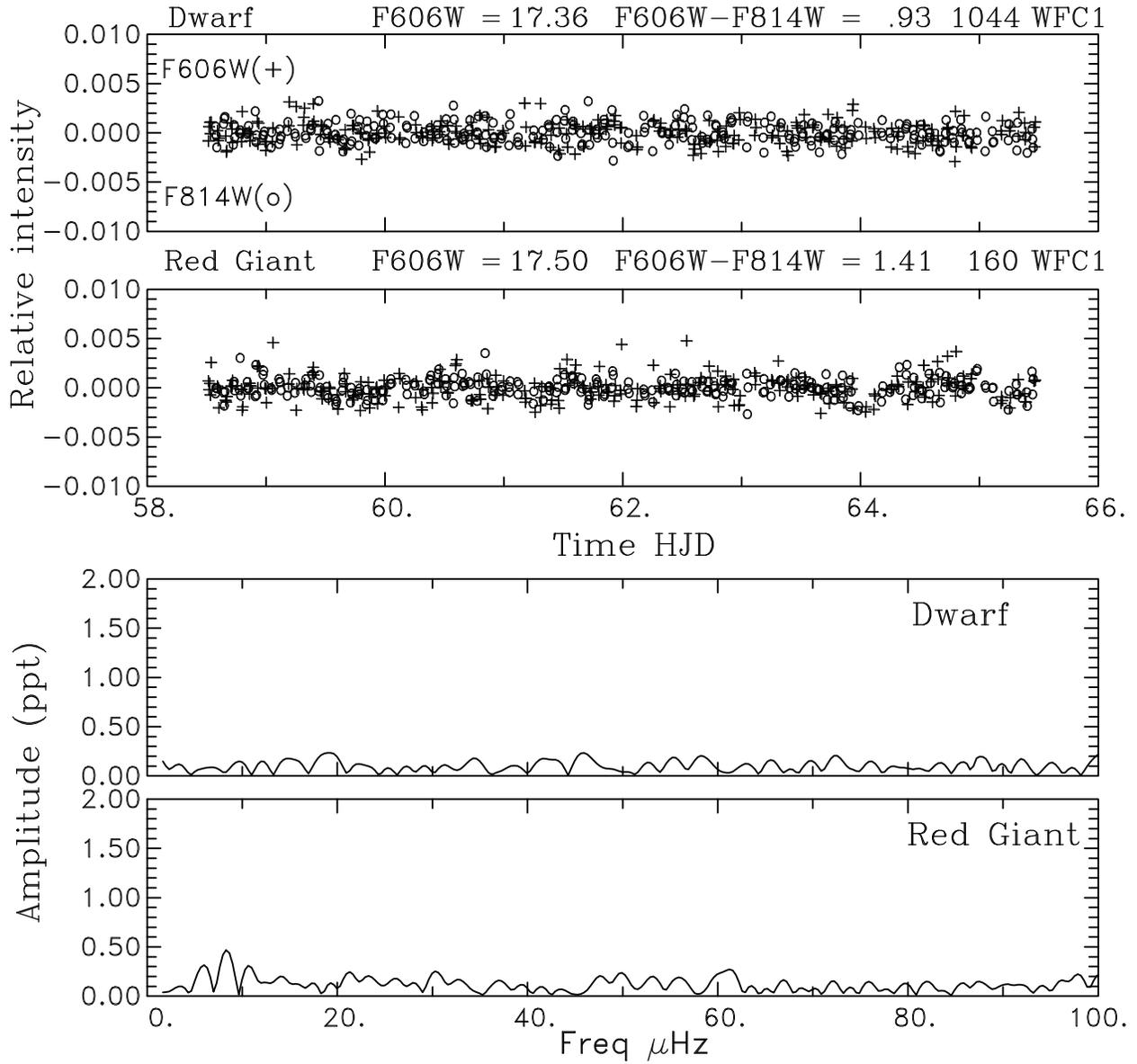}
\caption{Same as Figure 4, but for F606W $\sim$17.5.
\label{fig8}}
\end{figure}

\begin{figure}
\plotone{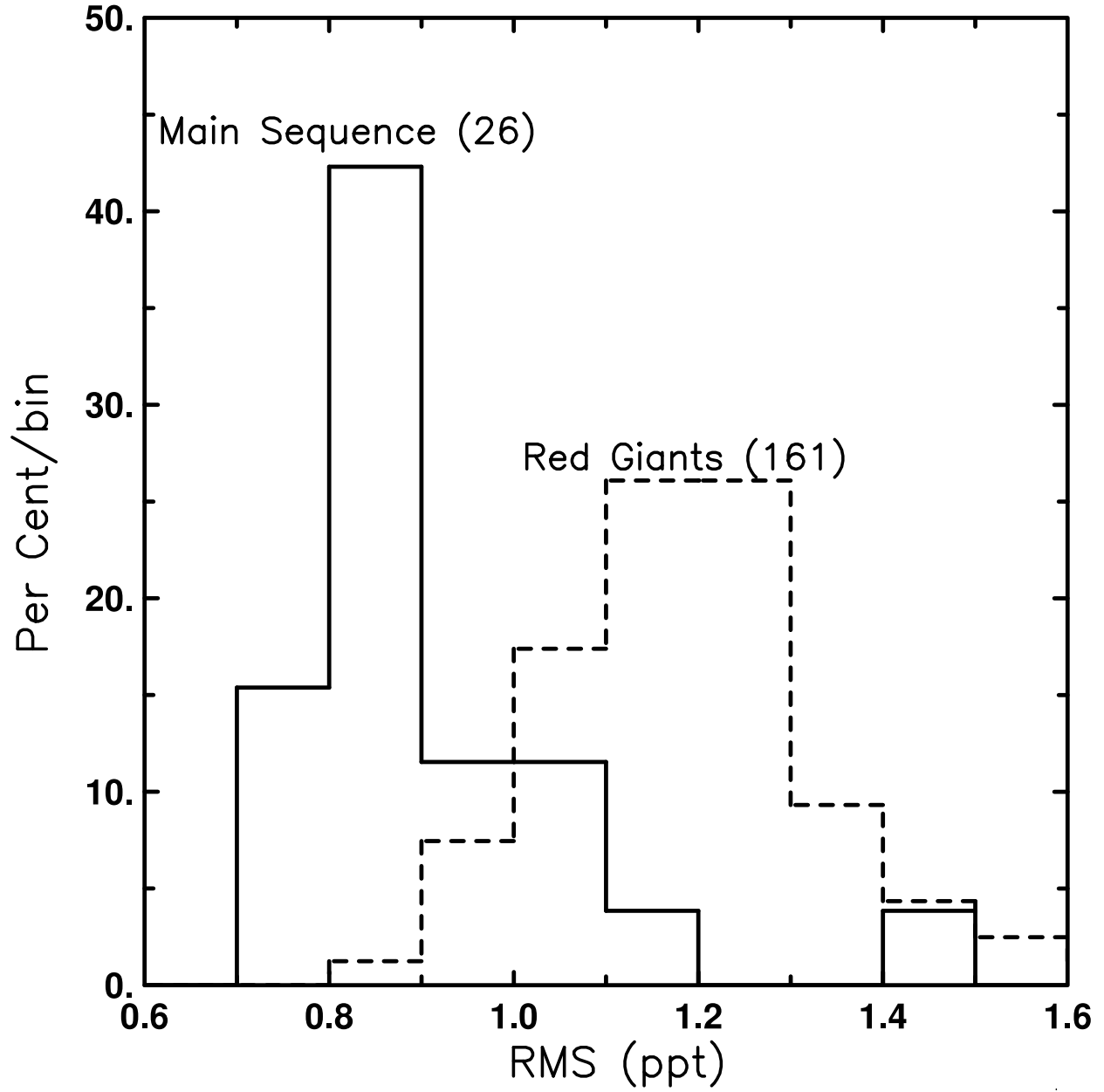}
\caption{Histogram of {\em rms} values for the F606W time series for dwarfs centered at F606W~= 16.5 and red giants at the same magnitude pulled from bins~{\bf B} and {\bf C}.
\label{fig9}}
\end{figure}

\begin{figure}
\plotone{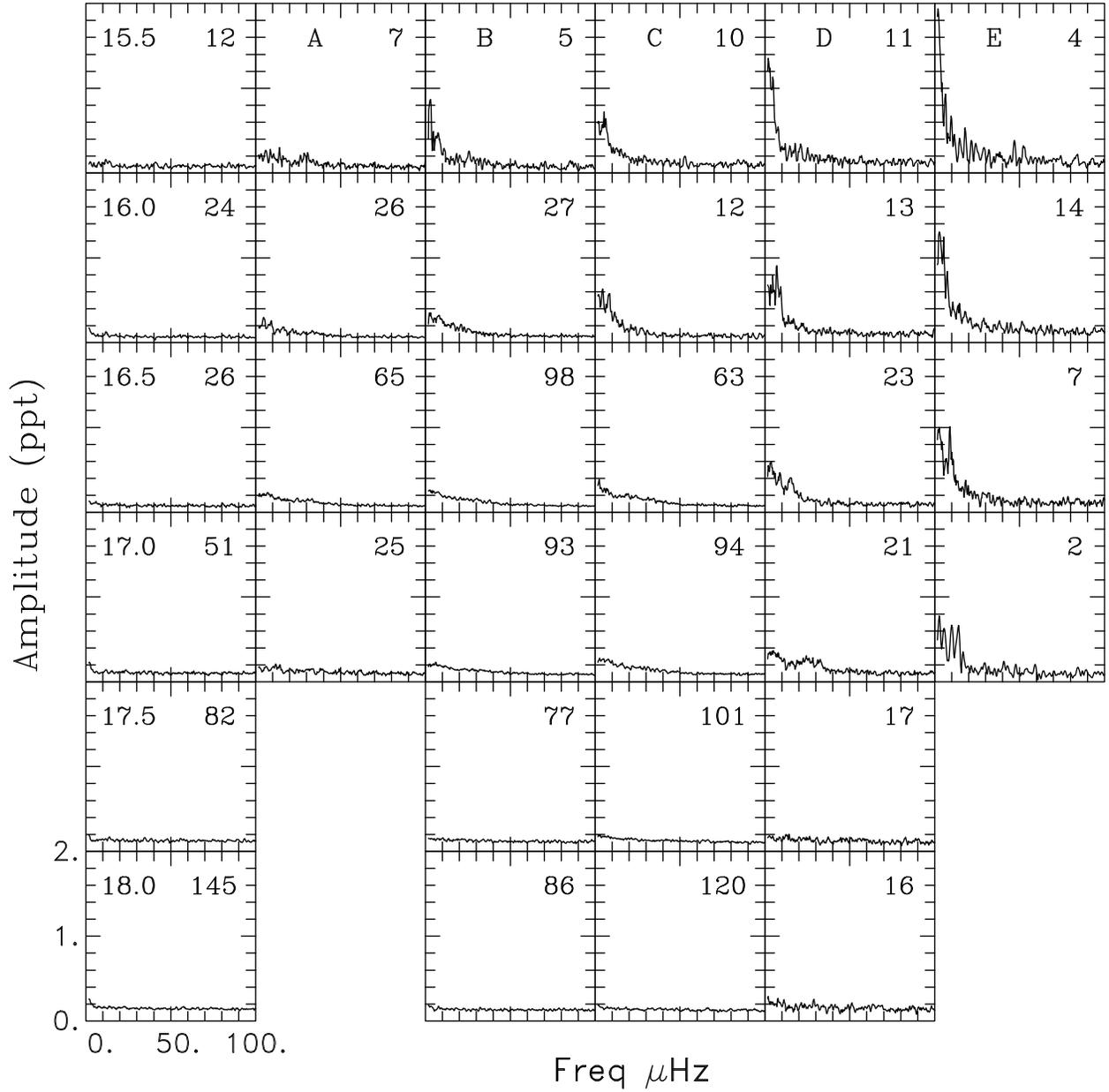}
\caption{Individual boxes map to the distribution shown in Figure~2 and Table~1, the frequency range and amplitude spectrum range are the same for each as shown for lower left case.  The number of stars contributing to the median amplitude spectrum is shown in the upper right of each box. The left column is for the dwarf star control sample.
\label{fig10}}
\end{figure}

\begin{figure}
\plotone{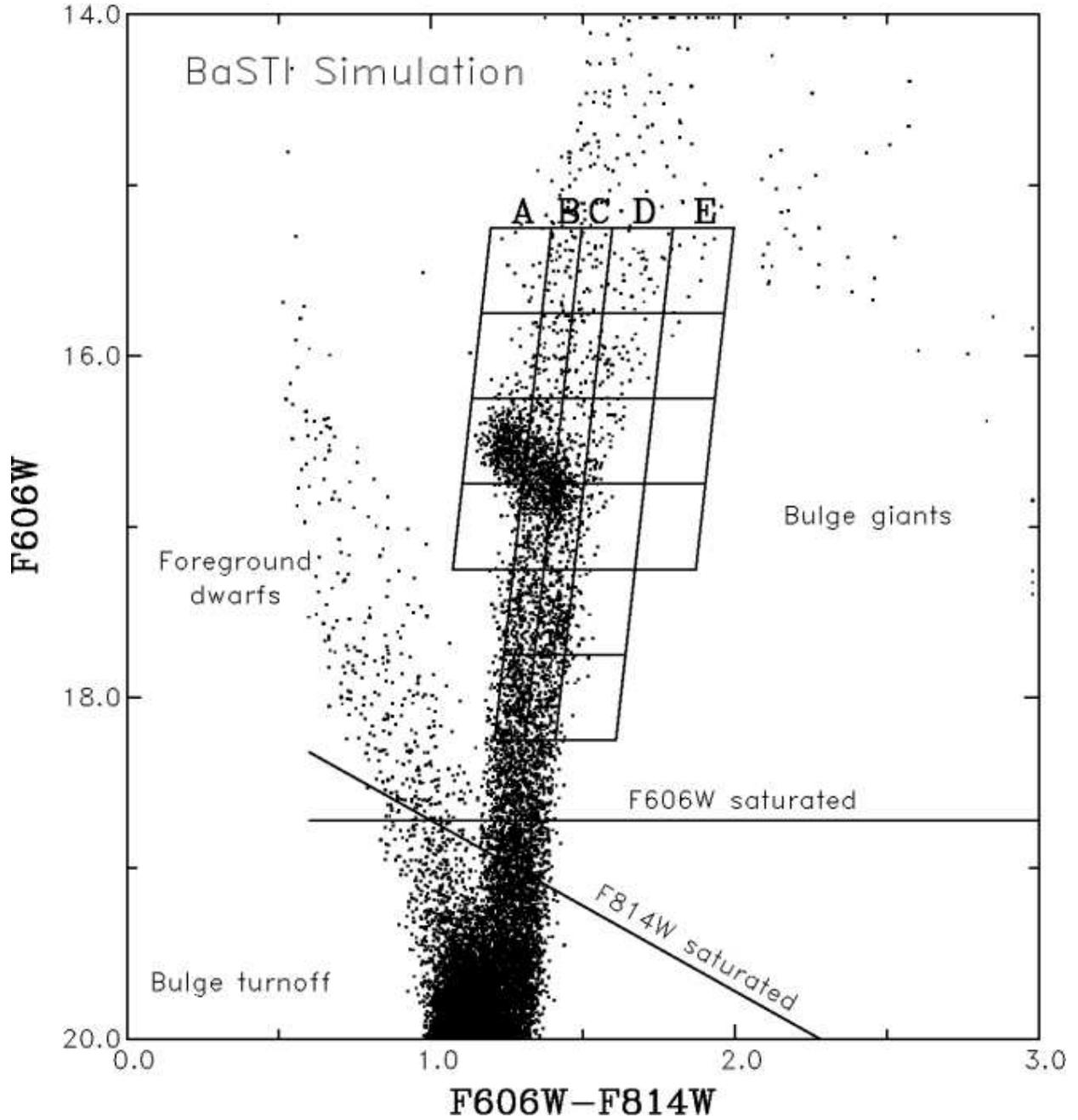}
\caption{Same as Figure 2, but with simulated stellar properties from the BaSTI codes.  See text for details. 
\label{fig11}}
\end{figure}

\begin{figure}
\plotone{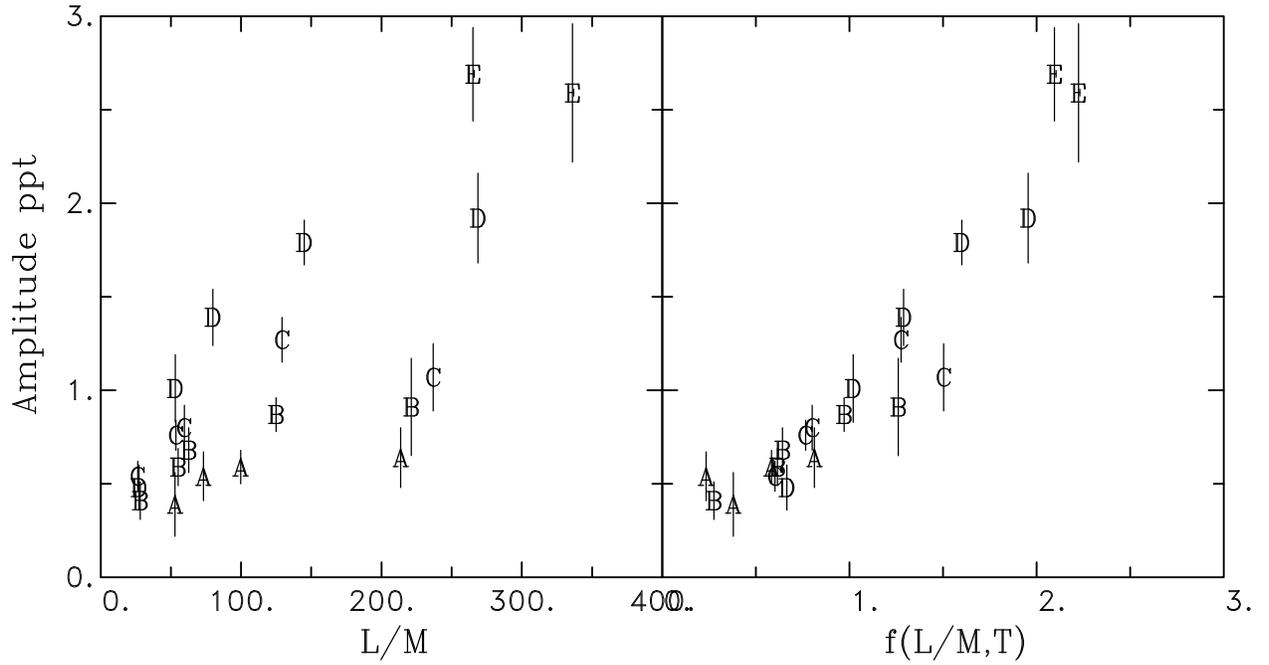}
\caption{Symbols and associated error bars correspond to the variation excess {\em rms} in parts per thousand as given in Table~1.  In the left panel these are plotted against the luminosity to mass (in solar units) ratio developed from
the analyses discussed in \S5.1.  The right panel plots the same data against a least squares regression in $L/M$ and $T_\mathrm{eff}$ as discussed in the text and developed in functional form as Eq.~(5).
\label{fig12}}
\end{figure}

\clearpage

\begin{deluxetable}{ccrrcc}
\tablewidth{0pt}
\tablecaption{Excess {\em rms} of Red Giants dependence on CMD position.\label{tbl-1}}
\tablehead{
\colhead{F606W} &\colhead{A} &\colhead{B} &\colhead{C} &\colhead{D} &\colhead{E}}
\startdata
15.5 & 0.64(.16) & 0.91(.26) & 1.07(.18) & 1.92(.24) & 2.59(.37) \\
16.0 & 0.59(.09) & 0.87(.09) & 1.27(.12) & 1.79(.12) & 2.69(.25) \\
16.5 & 0.54(.13) & 0.68(.12) & 0.80(.12) & 1.39(.15) & 1.90(.11) \\
17.0 & 0.39(.17) & 0.59(.10) & 0.76(.08) & 1.01(.18) & 1.34(.14) \\
17.5 & \nodata  & 0.41(.10) & 0.54(.08) & 0.48(.12) &  \nodata  \\
18.0 &  \nodata & $-$0.11(.20) & $-$0.10(.15) & 0.71(.21) & \nodata \\
\enddata
\tablecomments{See Figure~2 for correspondence of bins in table with position in the CMD.  From A to E spans lower to higher F606W--F814W color, or higher to lower temperatures.  Units are parts-per-thousand of measured {\em rms} excess of giants relative to dwarfs at the same F606W, errors on the mean in each bin are shown in parentheses.}
\end{deluxetable}

\begin{deluxetable}{clllll}
\tablewidth{0pt}
\tablecaption{Median of mean frequency -- dependence on CMD 
position.\label{tbl-2}}
\tablehead{
\colhead{F606W} &\colhead{A} &\colhead{B} &\colhead{C} &\colhead{D} &\colhead{E}} 
\startdata
15.5 & 17.7(5) & 11.05(4) & \phn7.2(8) & \phn4.0(6) & 5.0(2)\\
16.0 & 15.4(21) & 12.2(21) & \phn8.4(10) & \phn7.2(9) & 5.2(10) \\
16.5 & 18.9(43) & 17.5(70) & 16.3(43) & \phn9.2(18) & 6.7(5) \\
17.0 & 25.2(9) & 18.4(41) & 19.2(58) & 16.6(12) & \multicolumn{1}{c}{\nodata} \\
17.5 & \multicolumn{1}{c}{\nodata}  & 31.6(4) & 23.3(17) & 25.3(2) & \multicolumn{1}{c}{\nodata} \\
\enddata
\tablecomments{Values of the median frequency in amplitude spectra for each CMD bin with at least 2 stars having significance of highest peak up to 70~$\mu$Hz $<$ 0.001 \citep{sca82}.  The number of contributing stars is shown in parenthesis.  Units are $\mu$Hz.}
\end{deluxetable}

\begin{deluxetable}{clllll}
\tablewidth{0pt}
\tablecaption{Relative amplitude of F606W to F814W -- dependence on CMD 
position.\label{tbl-3}}
\tablehead{
\colhead{F606W} &\colhead{A} &\colhead{B} &\colhead{C} &\colhead{D} &\colhead{E}} 
\startdata
15.5 & 0.96(0.12,3) & 0.99(0.19,4) & 1.14(0.23,8) & 1.40(0.42,6) & 1.65(0.18,2) \\
16.0 & 0.97(0.18,15) & 1.07(0.12,20) & 1.33(0.16,10) & 1.37(0.24,9) & 1.72(0.22,8) \\
16.5 & 0.93(0.19,30) & 1.01(0.18,61) & 1.11(0.25,37) & 1.44(0.20,18) & 1.73(0.23,5) \\
17.0 & 1.14(0.10,5) & 0.96(0.19,33) & 1.08(0.19,48) & 1.14(0.19,11) & \multicolumn{1}{c}{\nodata} \\
17.5 & \multicolumn{1}{c}{\nodata} & \multicolumn{1}{c}{\nodata} & 0.96(0.16,10) & 1.07(0.05,2) & \multicolumn{1}{c}{\nodata} \\
\enddata
\tablecomments{Relative amplitude of F606W time series variations to same in F814W, see text for details.  Standard deviation and number of stars included in each bin are shown in parentheses.}
\end{deluxetable}

\begin{deluxetable}{cccccc}
\tablewidth{0pt}
\tablecaption{Mode multiplicity -- dependence on CMD position.\label{tbl-4}}
\tablehead{
\colhead{F606W} &\colhead{A} &\colhead{B} &\colhead{C} &\colhead{D} &\colhead{E}} 
\startdata
15.5 & 4.40 & 5.00 & 3.75 & 2.50 & 4.50 \\
16.0 & 3.48 & 4.45 & 5.00 & 4.22 & 3.40 \\
16.5 & 2.77 & 3.53 & 4.35 & 4.00 & 4.60 \\
17.0 & 2.78 & 2.66 & 3.12 & 4.33 & 5.00 \\
17.5 & \nodata & 1.25 & 1.82 & 1.00 & \nodata \\
\enddata
\tablecomments{Mean number of significant peaks per power spectrum---see
text for details.}
\end{deluxetable}

\begin{deluxetable}{crrrrc}
\tablewidth{0pt}
\tablecaption{Stellar Characteristics over CMD -- BaSTI simulation.\label{tbl-5}}
\tablehead{
\colhead{F606W} &\colhead{A} &\colhead{B} &\colhead{C} &\colhead{D} &\colhead{E} \\
\noalign{\hrule}
\multicolumn{6}{c}{Mean Luminosity (solar units)} } 
\startdata
15.5 & 164.6 & 190.1 & 217.8 & 270.4 & 350.3 \\  
16.0 &  84.2 & 109.1 & 123.3 & 148.6 & 277.3 \\  
16.5 &  57.7 &  57.1 &  58.2 &  84.5 & \nodata \\  
17.0 &  45.7 &  51.1 &  52.5 &  55.4 & \nodata \\  
17.5 &  25.9 &  25.0 &  27.2 &  28.8 & \nodata \\  
18.0 &  14.7 &  15.2 &  16.6 &  21.8 & \nodata \\  
\cutinhead{Effective Temperature (Kelvins)} 
15.5 & 4887 & 4550 & 4385 & 4093 & 3949 \\  
16.0 & 4943 & 4677 & 4457 & 4246 & 3999 \\  
16.5 & 5176 & 4819 & 4688 & 4375 & \nodata \\  
17.0 & 5000 & 4819 & 4699 & 4508 & \nodata \\  
17.5 & 5058 & 4977 & 4710 & 4667 & \nodata \\  
18.0 & 5135 & 5093 & 4819 & 4732 & \nodata \\  
\cutinhead{Mean Stellar Radii (solar units)} 
15.5 & 17.97 & 22.28 & 25.66 & 32.83 & 40.14 \\  
16.0 & 12.56 & 15.97 & 18.70 & 22.61 & 34.82 \\  
16.5 &  9.48 & 10.88 & 11.61 & 16.06 & \nodata \\  
17.0 &  9.04 & 10.29 & 10.97 & 12.25 & \nodata \\  
17.5 &  6.65 &  6.75 &  7.87 &  8.24 & \nodata \\  
18.0 &  4.86 &  5.02 &  5.86 &  6.98 & \nodata \\  
\enddata
\tablecomments{See text for discussion of simulations based on BaSTI codes.}
\end{deluxetable}
\clearpage

\begin{deluxetable}{crrrrc}
\tablewidth{0pt}
\tabletypesize{\footnotesize}
\tablecaption{Predicted frequency characteristics.
\label{tbl-6}}
\tablehead{
\colhead{F606W} &\colhead{A} &\colhead{B} &\colhead{C} &\colhead{D} &\colhead{E} \\
\noalign{\hrule}
\multicolumn{6}{c}{Frequency of maximum power ($\mu$Hz)} }
\startdata
15.5 &   7.83 &   5.89 &   4.84 &   3.35 &   2.36 \\  
16.0 &  17.48 &  11.49 &   9.39 &   7.07 &   3.13 \\  
16.5 &  28.00 &  25.45 &  24.34 &  14.23 & \nodata \\  
17.0 &  34.27 &  28.88 &  27.24 &  23.81 & \nodata \\  
17.5 &  63.05 &  63.35 &  55.71 &  52.39 & \nodata \\  
18.0 & 117.53 & 112.16 &  99.17 &  71.89 & \nodata \\  
\cutinhead{Frequency of 3rd radial overtone ($\mu$Hz)} 
15.5 &   8.11 &   6.13 &   5.20 &   3.74 &   2.80 \\  
16.0 &  14.45 &  10.29 &   8.42 &   6.65 &   3.54 \\  
16.5 &  21.32 &  18.61 &  17.57 &  11.23 & \nodata \\  
17.0 &  23.92 &  20.38 &  19.13 &  16.74 & \nodata \\  
17.5 &  38.06 &  37.64 &  32.23 &  30.46 & \nodata \\  
18.0 &  60.93 &  58.44 &  50.22 &  38.99 & \nodata \\  
\cutinhead{Large frequency splitting ($\mu$Hz)} 
15.5 &   .78 &   .59 &   .50 &   .36 &   .27 \\  
16.0 &  1.39 &   .99 &   .81 &   .64 &   .34 \\  
16.5 &  2.05 &  1.79 &  1.69 &  1.08 & \nodata \\  
17.0 &  2.30 &  1.96 &  1.84 &  1.61 & \nodata \\  
17.5 &  3.66 &  3.62 &  3.10 &  2.93 & \nodata \\  
18.0 &  5.86 &  5.62 &  4.83 &  3.75 & \nodata \\ 
\enddata 
\tablecomments{All quantities use the BaSTI derived stellar parameters as shown in Table~5. Evaluation of the predicted location of maximum oscillation power uses Eq.~(1). Scaling of 3rd overtone frequency \citep{kal05} by $\rho^{1/2}$  provides an indication of low-order radial mode frequencies. Prediction for separation of modes in frequency space as half of the large splitting is from Eq.~(2).}
\end{deluxetable}
\clearpage

\begin{deluxetable}{crrrrc}
\tablewidth{0pt}
\tablecaption{Predicted oscillation amplitudes.
\label{tbl-7}}
\tablehead{
\colhead{F606W} &\colhead{A} &\colhead{B} &\colhead{C} &\colhead{D} &\colhead{E} \\
\noalign{\hrule} 
\multicolumn{6}{c}{\citet{kje95} -- Eq.~(3)}}
\startdata
15.5 &  1.28 &  1.52 &  1.76 &  2.29 &  3.07 \\  
16.0 &   .58 &   .81 &   .93 &  1.15 &  2.36 \\  
16.5 &   .39 &   .38 &   .39 &   .59 & \nodata \\  
17.0 &   .30 &   .34 &   .35 &   .37 & \nodata \\  
17.5 &   .17 &   .16 &   .17 &   .18 & \nodata \\  
18.0 &   .09 &   .09 &   .10 &   .13 & \nodata \\  
\cutinhead{\citet{sam07} -- Eq.~(4)} 
15.5 &   .18 &   .19 &   .20 &   .21 &  \phn.25 \\  
16.0 &   .11 &   .13 &   .13 &   .14 &   \phn.21 \\  
16.5 &   .09 &   .08 &   .07 &   .09 & \nodata \\  
17.0 &   .07 &   .07 &   .07 &   .07 & \nodata \\  
17.5 &   .05 &   .04 &   .04 &   .04 & \nodata \\  
18.0 &   .03 &   .03 &   .03 &   .04 & \nodata \\  
\enddata
\tablecomments{Amplitudes in parts-per-thousand for comparison with observed {\em rms} given in Table 1.}
\end{deluxetable}

\end{document}